\documentclass[%
  aps,
  reprint,
  a4paper,
  twocolumn,
  superscriptaddress,
  preprintnumbers,
  nofootinbib,
  prd
]{revtex4-1}

\usepackage{graphicx,color}
\usepackage{amsmath,amssymb,amsfonts}
\usepackage{stackrel}
\usepackage{enumitem}
\usepackage[english]{babel}
\usepackage{verbatim}
\usepackage{hyperref}
\usepackage{comment}
\usepackage{ulem}
\usepackage{subfigure}

\graphicspath{
  {./}
  {figures/}
}

\newcommand{\msbar}{{\overline{\mbox{\rm MS}}}}
\newcommand\he[1]{#1^\dagger} % Hermitian conjugate

\newcommand\td{\text{3D}}% index notation for 3d fields
\newcommand\fd{\text{4D}}% index notation for 4d fields

\newcommand\RE{\operatorname{Re}}

\begin{document}

%% addresses

\newcommand{\NTNU}{\affiliation{Department of Physics, Faculty of
    Natural Sciences, Norwegian University of Science and Technology, \\
    H{\o}gskoleringen 5, N-7491 Trondheim, Norway}}

\newcommand{\HIPetc}{\affiliation{ Department of Physics and Helsinki
    Institute of Physics, P.O. Box 64, FI-00014 University of Helsinki,
    Finland }}

\newcommand{\NBI}{\affiliation{Niels Bohr Institute, University of
    Copenhagen, Blegdamsvej 17, 2100 Copenhagen}}

\newcommand{\UiS}{\affiliation{Department of Mathematics and Physics,
  University of Stavanger, 4036 Stavanger, Norway}}

\newcommand{\Virginia}{\affiliation{Department of Physics, University
    of Virginia, 382 McCormick Road, Charlottesville, Virginia
    22904-4714, U.S.A.}}

\title{Nonperturbative Analysis of the Electroweak Phase Transition
  \\ in the Two Higgs Doublet Model}

\preprint{HIP-2017-26/TH}

\author{Jens~O.~Andersen} \email{andersen@tf.phys.ntnu.no} \NTNU

\author{Tyler~Gorda} \email{tyler.gorda@virginia.edu} \HIPetc \Virginia

\author{Andreas Helset} \email{ahelset@nbi.ku.dk} \NTNU \NBI

\author{Lauri~Niemi} \email{lauri.b.niemi@helsinki.fi} \HIPetc

\author{Tuomas~V.~I.~Tenkanen} \email{tuomas.tenkanen@helsinki.fi} \HIPetc

\author{Anders~Tranberg} \email{anders.tranberg@uis.no} \UiS

\author{Aleksi~Vuorinen} \email{aleksi.vuorinen@helsinki.fi} \HIPetc

\author{David~J.~Weir} \email{david.weir@helsinki.fi} \HIPetc

\begin{abstract}
We perform a nonperturbative study of the electroweak phase
transition (EWPT) in the two Higgs doublet model (2HDM) by deriving a
dimensionally reduced high-temperature effective theory for the model,
and matching to known results for the phase diagram of the effective
theory. We find regions of the parameter space where the theory
exhibits a first-order phase transition.  In particular, our findings
are consistent with previous perturbative results suggesting that the
primary signature of a first-order EWPT in the 2HDM is $m_{A_0} >
m_{H_0} + m_Z$.
\end{abstract}

\maketitle

\section{Introduction}

Accounting for the baryon asymmetry in the
present universe is a major unsolved problem in cosmology. One of the
leading candidates for a viable mechanism, electroweak baryogenesis
(EWBG) \cite{Kuzmin:1985mm}, suggests that the asymmetry originates
from the electroweak phase transition (EWPT) in the early
universe. According to the Sakharov conditions \cite{Sakharov:1967dj}
the transition would have to be first order, accompanied by a sizable
violation of CP-symmetry. Unfortunately, these conditions immediately
rule out EWBG within the minimal Standard Model (SM), as it was
demonstrated that the SM EWPT is a crossover
\cite{Kajantie:1995kf,Kajantie:1996mn,Kajantie:1996qd,Csikor:1998eu},
and that SM CP-violating effects are heavily suppressed at high
temperatures~\cite{Gavela:1994dt,Brauner:2011vb,Brauner:2012gu}.

Independently of the question of baryon asymmetry, a host of beyond
the Standard Model (BSM) theories have been proposed to solve open
problems in physics. Determining whether BSM theories can produce a
first order EWPT and thus facilitate EWBG is nontrivial:
quantitatively reliable conclusions about the phase transition
typically require a non-perturbative approach, deemed unmanageable for
large parameter spaces. Because of this difficulty, analyses based on
the finite-temperature effective potential have become
standard~\cite{Carrington:1991hz,Fromme:2006cm,Cline:2011mm,Dorsch:2013wja,Haarr:2016qzq,Alanne:2016wtx,Marzola:2017jzl}.
Such studies can, however, have considerable uncertainties,
particularly for physical observables: in one
study~\cite{Laine:2012jy}, errors in excess of 10\% in the critical
temperature and 50\% in the latent heat were found, compared to
non-perturbative studies.

In contrast, a more reliable approach uses dimensionally reduced
effective theories, originally applied to the SM in
Refs.~\cite{Kajantie:1995dw,Kajantie:1995kf,Kajantie:1996mn,Kajantie:1996qd,Laine:1998qk},
and recently applied to the SM accompanied by a real
singlet~\cite{Brauner:2016fla}. In this paper, we use this method to
treat a widely studied BSM model, the two Higgs doublet model (2HDM),
where the SM is augmented with an additional Higgs doublet [see
Ref.~\cite{Branco:2011iw} for a review, and
Refs.~\cite{Losada:1996ju,Losada:1998at,Andersen:1998br} for earlier
work on dimensional reduction (DR) in the 2HDM]. We derive a three-dimensional high-$T$
effective theory, studying regions of parameter space where this
theory has the same form as that of the Standard Model, similar to
Ref.~\cite{Cline:1996cr}. This reduces determining the phase diagram
of the theory to mapping its parameter space to that of the SM
effective theory. Equipped with the analysis of
\cite{Kajantie:1995kf,Kajantie:1996mn,Kajantie:1996qd}, we discover
interesting and phenomenologically viable regions of parameter space
where the EWPT is first order, corroborating key findings of
perturbative studies of EWBG in the 2HDM.

\section{Dimensional reduction \\ of the 2HDM}

Our four-dimensional
starting theory can be described by the schematic action
\begin{equation}
S=\int d^4x\, \left[\mathcal{L}_\text{gauge} +
  \mathcal{L}_\text{fermion} + \mathcal{L}_\text{scalar} +
  \mathcal{L}_\text{Yukawa}
\right], 
\end{equation}
suppressing counterterm and ghost contributions. The field content
includes $\mathrm{SU}(3)_c$, $\mathrm{SU}(2)_L$ and $\mathrm{U}(1)_Y$
gauge fields, two scalar doublets $\phi_1$ and $\phi_2$, as well as
all fermions present in the SM. In our present treatment, we will
consider only one quark flavor in the Yukawa sector, namely the top,
since it has the largest coupling to the Higgs field. The top quark
couples to one doublet only (by convention $\phi_2$), and we have not
yet committed to a specific type of 2HDM (I or II).

The extended scalar sector of our model reads
\begin{equation}
\mathcal{L}_\text{scalar}=\sum_{i=1}^2(D_\mu\phi_i)^\dagger (D_\mu\phi_i) + V(\phi_1,\phi_2), \label{eq:4D}
\end{equation}
with usual covariant derivative $D_\mu$ and the potential
\begin{multline}
V(\phi_1,\phi_2)  =  \mu^2_{11}\phi_1^\dagger \phi_1 + \mu^2_{22} \phi_2^\dagger \phi_2 + \mu^2_{12} \phi_1^\dagger \phi_2 + \mu^{2*}_{12} \phi_2^\dagger \phi_1 \\
  + \lambda_1 (\phi_1^\dagger \phi_1)^2 + \lambda_2 (\phi_2^\dagger \phi_2)^2 + \lambda_3 (\phi_1^\dagger \phi_1)(\phi_2^\dagger \phi_2) \\ + \lambda_4 (\phi_1^\dagger \phi_2)(\phi_2^\dagger \phi_1)  
 + \frac{\lambda_5}{2} (\phi_1^\dagger \phi_2)^2  +
 \frac{\lambda^*_5}{2} (\phi_2^\dagger \phi_1)^2. \\
\end{multline}
In general, C(P) symmetry is broken when $\lambda_5$ or $\mu_{12}^2$
are complex; we have discarded so-called hard CP-breaking terms, often
parametrised by $\lambda_{6,7}$, cf.~\cite{Branco:2011iw,Gorda:2018hvi}.

The first three-dimensional effective theory, obtained by integrating
out the `superheavy' hard scale $\pi T$ (see
e.g.~Ref.~\cite{Brauner:2016fla} for details of the
procedure), has schematic form
\begin{equation}
\label{eq:full_3d_lag}
S = \int d^3x\, \left[\mathcal{L}^{(3)}_\text{gauge} +
  \mathcal{L}^{(3)}_\text{scalar} + \mathcal{L}^{(3)}_\text{temporal}
\right],
\end{equation}
again suppressing ghost and counterterm contributions. The field
content is now $\mathrm{SU}(2)_L$ and $\mathrm{U}(1)_Y$ gauge fields;
two Higgs doublets; and temporal scalar fields $A^a_0, B_0,
C^\alpha_0$. The fermions are integrated out and the
$\mathrm{SU}(3)_c$ gauge fields can be
neglected~\cite{Brauner:2016fla}.  The fundamental scalar sector
remains of the form
\begin{equation}
\mathcal{L}^{(3)}_\text{scalar}=(D_r\phi_1)^\dagger (D_r\phi_1) +  (D_r\phi_2)^\dagger (D_r\phi_2)+V(\phi_1,\phi_2),\nonumber\\
\end{equation}
where $r=1,2,3$ is summed over. In the second step of DR, the
heavy temporal scalar fields are integrated out.

Although the theory in (\ref{eq:full_3d_lag}) is already suitable for
lattice simulations, it can be further simplified by noticing that
$\phi_1$ and $\phi_2$ mix when $\mu^2_{12}\neq 0$, and near the phase
transition there typically exists a hierarchy between the mass
eigenvalues. This observation---specific to the 2HDM---allows us to
integrate out the heavy mode and study the phase transition with only
one scalar field coupled to the gauge fields. Our final effective
theory becomes
\begin{align}
\label{eq:final_3d_lag}
S &= \int d^3x\, \left[\hat{\mathcal{L}}^{(3)}_\text{gauge} +
  \hat{\mathcal{L}}^{(3)}_\text{scalar}
\right], \\
\label{eq:3D}
\hat{\mathcal{L}}^{(3)}_\text{scalar} &= (D_r\phi)^\dagger (D_r\phi) +
\hat{\mu}^2_3 \phi^\dagger \phi + \hat{\lambda}_3 (\phi^\dagger
\phi)^2.
\end{align}
Here, $\phi$ is the remaining light $\phi_1$-$\phi_2$ mode, and
the parameters of the theory include $\hat{\mu}^2_3$, $\hat{\lambda}_3$
and the 3D gauge couplings $\hat{g}_3'$ and $\hat{g}_3$ for the
$\mathrm{U}(1)_Y$ and $\mathrm{SU}(2)_L$ interactions. As in the analysis of
Refs.~\cite{Kajantie:1995dw,Kajantie:1995kf}, we omit all
non-perturbative effects related to the $\mathrm{U}(1)_Y$ field.

The main task of DR is to perturbatively match the parameters of the
original 4D theory, Eq.~(\ref{eq:4D}), to those of the final effective
theory, Eq.~(\ref{eq:3D}). This is accomplished by demanding that the
effective theory reproduces the static Green's functions of the
original theory at large distances $R \gg 1/T$. This results in a
number of matching relations from which the effective theory
parameters are solved. This procedure is presented in
Ref.~\cite{Gorda:2018hvi} and summarised in the Supplemental
Material~\cite{supplemental}.

As discussed above, the effective theory of Eq.~(\ref{eq:3D}) has the
\textsl{same} form as that of the SM, studied in
Refs.~\cite{Kajantie:1995kf,Kajantie:1996mn,Kajantie:1996qd}, but with
\textsl{different} matching relations. This allows us to adopt
existing numerical results for the strength of the phase transition,
and study the phase diagram through our matching procedure alone.

The validity of DR can be quantified by estimating the effect of
neglected dimension-six operators. While it is difficult to
comprehensively gauge their effect, one can evaluate the change in the
vacuum expectation value (vev) of the Higgs field in the effective
theory caused by the $(\phi^\dagger \phi)^3$ operators. In Eq.~(201)
of Ref.~\cite{Kajantie:1995dw}, it was shown that in the SM the
dominant neglected contribution comes from the top quark; its effect
is about one percent. In the first DR step where the superheavy fields
are integrated out, we estimate the effect of new BSM contributions by
comparing their magnitude to the contribution from the top quark [see
  Eqs.~(\ref{eq:dim6op1},\ref{eq:dim6op2}) in the Supplemental
  Material].  However, in many cases the operator $\mathcal{O}^{(6)}_B
= \hat{\Lambda}_6 (\phi^\dagger\phi)_{\td}^3$ generated when the heavier doublet
is integrated out dominates over the six-dimensional operators of the
first step, denoted $\{\mathcal{O}^{(6)}_{A,i}\}$. We discuss
  these operators in detail below.

Finally, although the parameter matching is perturbative, the study of
the 3D phase diagram is non-perturbative and---within the limitations
of lattice methods---exact. The main advantage of our approach lies in
proper handling of the infrared physics, which causes trouble in
traditional perturbative studies of the EWPT. Resummations are
performed when the superheavy and heavy scales are integrated out
perturbatively, and the problematic light modes are treated
non-perturbatively on the lattice. However, the mapping to precise
values of the 4D parameters, where this phase transition occurs in
the 2HDM, is limited by the accuracy of the perturbative
truncation. We organise the expansion in terms of the gauge coupling
$g$, and perform the DR to $O(g^4)$. Thus the calculation is
carried out at the one-loop level for quartic couplings, and two-loop
level for mass parameters. This exceeds the accuracy used in the
perturbative calculations of e.g.~Ref.~\cite{Dorsch:2014qja} (see,
however, Ref.~\cite{Laine:2017hdk} for a recent two-loop perturbative
calculation in the inert doublet model).  The uncertainty in the
effective theory due to the choice of renormalisation scale is
discussed in the Supplemental Material.

\section{Scanning the parameter space}

The phase diagram of the
dimensionally-reduced theory can be mapped using the dimensionless
parameters $x\equiv \hat{\lambda}_3/\hat{g}^2_3$, $y\equiv
\hat{\mu}^2_3/\hat{g}^4_3$. It is known that within this theory the
EWPT occurs near $y\simeq 0$, where the Higgs mass parameter becomes
negative. In
Refs.~\cite{Kajantie:1995kf,Kajantie:1996mn,Kajantie:1996qd}, it was
found that the transition is first order for $x \lesssim 0.11$, and
strongly so for $x \lesssim 0.04$. In this paper we are focussed on
finding where the crossover turns into a first-order transition.

We search for areas of 2HDM parameter space that map onto regions of
the 3D effective theory with $x<0.11$ and $y\simeq 0$. Since there
are ten real parameters in the 4D theory and only three in the 3D
one, inverting the mapping process is not unique. We perform scans of
the 2HDM parameter space, guided by the results of
Ref.~\cite{Dorsch:2017nza} that combine phenomenological constraints
with a one-loop resummed perturbative determination of the effective
potential. Other recent treatments are found in
Refs.~\cite{Basler:2016obg,Basler:2017uxn}.

\begin{figure*}
\begin{center}
  \subfigure[\ $\tan(\beta) =
    2.0$]{\includegraphics[width=0.32\textwidth]{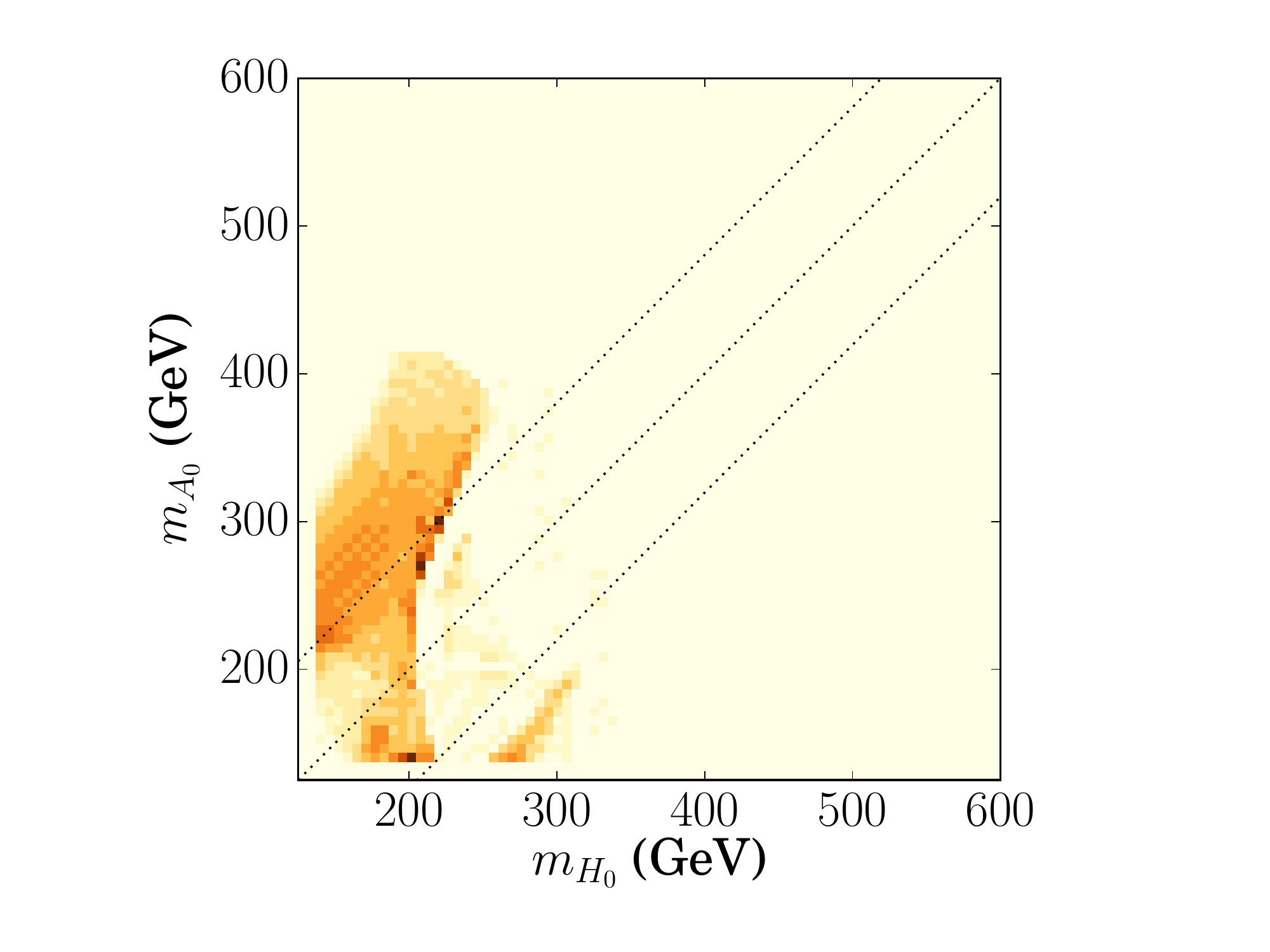}} 
  \subfigure[\ $\tan(\beta) =
    2.5$]{\includegraphics[width=0.32\textwidth]{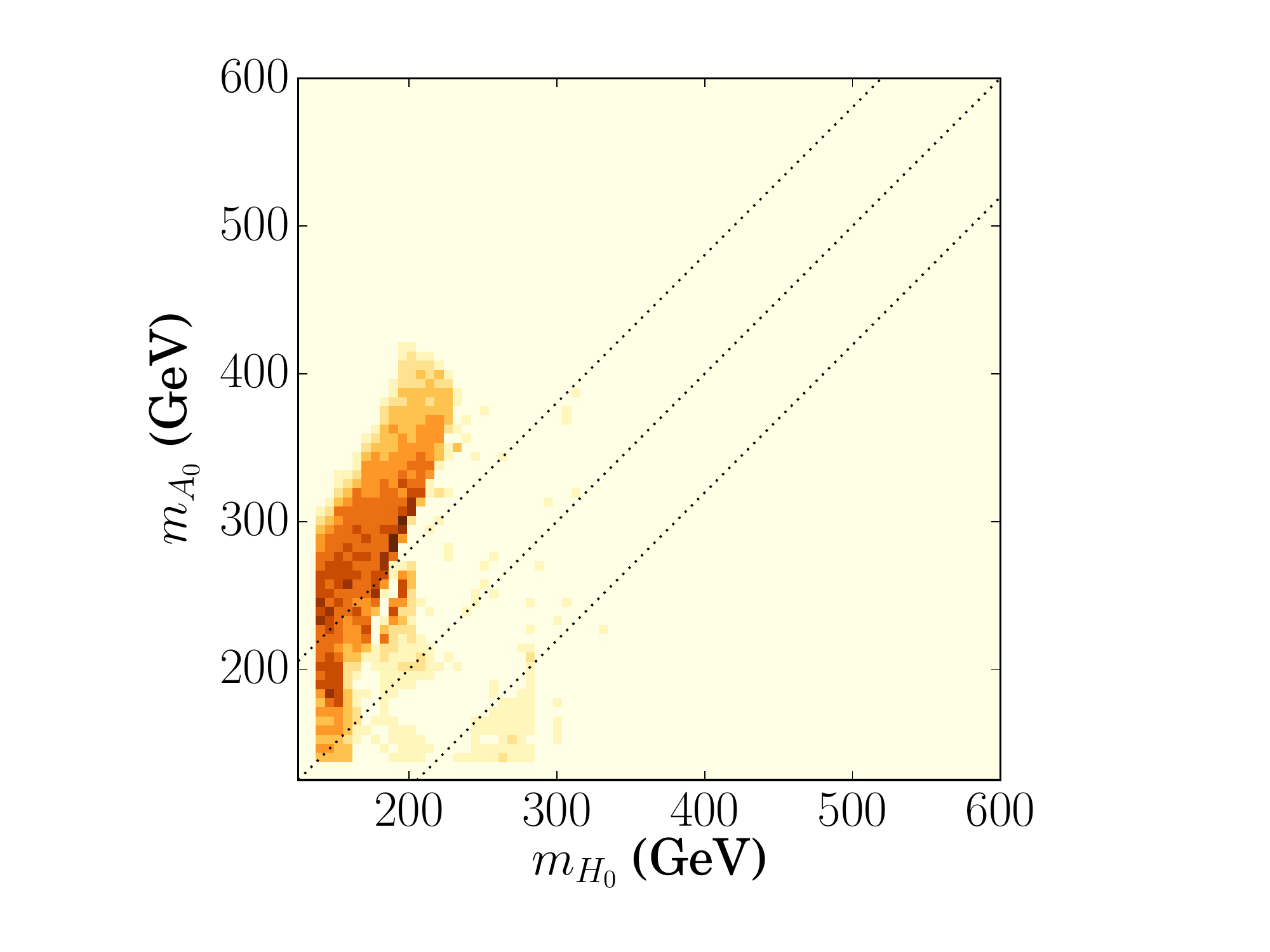}} 
  \subfigure[\ $\tan(\beta) =
    3.0$]{\includegraphics[width=0.32\textwidth]{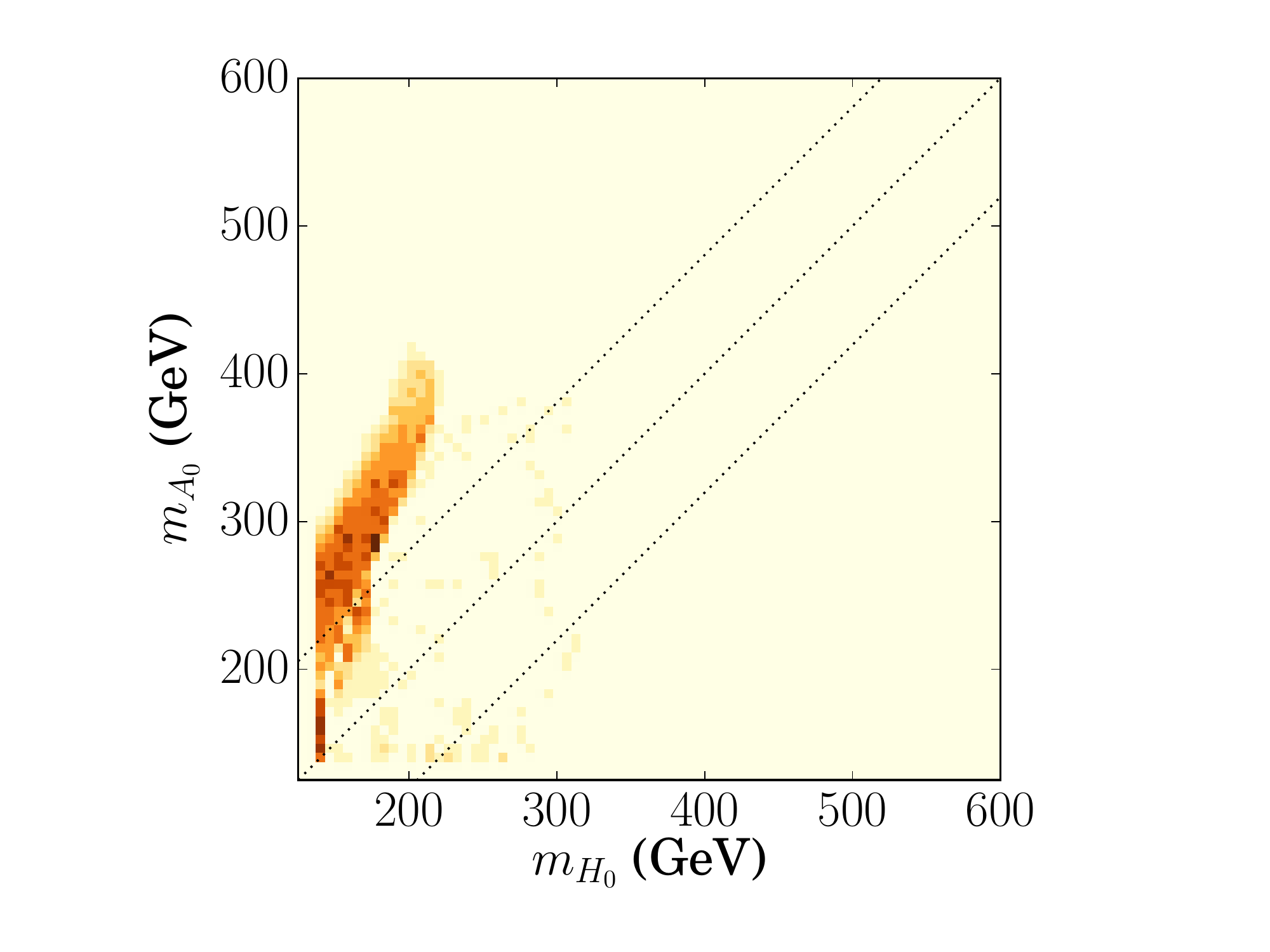}} 
\end{center}
\caption{\label{fig:separate_heatmaps} Heat maps with fixed
  $\tan(\beta)$, showing regions of first order EWPT ($0 < x < 0.11$
  and $y \simeq 0$) in the alignment limit. The dotted lines
  correspond to $m_{A_0} = m_{H_0}$ and $m_{A_0} = m_{H_0} \pm m_Z$.}
\end{figure*}

A uniform scan through a 10-dimensional space is computationally
expensive; we must therefore make some simplifying assumptions. We
take all parameters of the 2HDM to be real, setting
$\text{Im}(\lambda_5) = 0$, $\text{Im}(\mu^2_{12}) = 0$. This
eliminates extra CP violation in the model, which would be crucial for
baryogenesis. However, the effect of these imaginary parts on the
strength of the transition is expected to be negligible; the
CP-violating phase must necessarily be small due to EDM
constraints~\cite{Inoue:2014nva,Dorsch:2016nrg,Chen:2017com}.

Next, we reparametrise the model following Ref.~\cite{Dorsch:2017nza},
applying tree-level relations between the $\msbar$ parameters and
physical quantities; accounting for (possibly sizeable) loop effects
from vacuum renormalisation is left for future work.  The masses of
the CP-even scalars are denoted by $m_{h} = 125~\mathrm{GeV}$ and
$m_{H_0}$; that of the CP-odd scalar by $m_{A_0}$; and that of the
charged scalar by $m_{H^\pm}$. We also employ two angles $\alpha$ and
$\beta$: $\alpha$ parametrises mixing between the CP-even states,
while $\beta$ is related to the ratio of the vevs $\tan(\beta) \equiv
\frac{\nu_2}{\nu_1}$. Here, $\nu_1$ and $\nu_2$ are the vevs for
$\phi_1$ and $\phi_2$, respectively, with $\nu^2_1 + \nu^2_2 = \nu^2$
and $\nu = 246~\mathrm{GeV}$. Finally, there is the squared mass scale
$M^2\equiv \mu^2(\tan(\beta) + 1/\tan(\beta))$, where we treat $\mu^2
\equiv -\mathrm{Re}\, \mu^2_{12}$ as an input parameter. The relations
between the physical states and gauge eigenstates can be obtained from
Ref.~\cite{Dorsch:2017nza}.\footnote{In Eq.~(A.1) of
  Ref.~\cite{Dorsch:2017nza}, there is a misprint in the powers of
  $\tan(\beta)$ in the equations for $\lambda_1$ [$\cos(\beta-\alpha)
  \tan(\beta) \rightarrow \cos(\beta-\alpha) /\tan(\beta)$] and
  $\lambda_2$ [$\cos(\beta-\alpha)/\tan(\beta) \rightarrow
  \cos(\beta-\alpha) \tan(\beta)$].}

We also fix $m_{H^\pm} = m_{A_0}$, since EW precision tests require
the mass of the charged Higgs to be roughly degenerate with either
$H_0$ or $A_0$~\cite{Grimus:2007if,Grimus:2008nb}. Furthermore, we
work in the alignment limit, setting $\cos(\beta-\alpha) = 0$. In
this limit, the CP-even scalar $h$ couples to SM particles exactly
like the SM Higgs. We investigate relatively few values for
$\tan(\beta)$, whereas we perform a more exhaustive scan in a
three-dimensional parameter space spanned by $m_{H_0}$, $m_{A_0}$, and
$\mu^2$. At each point, we require that tree-level stability and
unitary constraints be satisfied; for details, see
Ref.~\cite{Gorda:2018hvi}. Furthermore, for the scaling assumptions of
DR to be valid, the tree-level mass parameters $\mu_{11}$, $\mu_{22}$
and $\mu_{12}$ should be comparable to the Debye mass $m_D \sim gT$
near the phase transition.  This sets an upper bound for the input
parameter $\mu \lesssim 200~\mathrm{GeV}$. Finally, we verify that in
the effective theory the other doublet really is heavy near the phase
transition, so it is justified to integrate it out.

\section{Results}

Following our scanning protocol outlined above, we
fix $\tan(\beta)$ and scan in the two scalar masses $m_{H_0}$ and
$m_{A_0}$ between $137.5$ and $562.5~\mathrm{GeV}$ at spacings of
$6.25~\mathrm{GeV}$, a total of 4624 points. A dense scan in $\mu$ is
then carried out for each pair, from 10 to 150 GeV at intervals of 2.5
GeV for a total of 56 values. In all, each of our fixed-$\tan(\beta)$
plots results from scanning approximately 260 000 points. The upper
limit on $\mu$ is chosen to ensure that DR is valid, as explained
above.

We first check whether each point is physical, according to our
criteria. If so, we then perform DR for evenly-spaced temperatures
between 80 and 200 GeV, at intervals of 20 GeV. This allows us to find
the value of $x$ when $y=0$---on the critical line---by
interpolation. We then use $x$ to characterise the phase transition.
We take $0.0 < x < 0.11$ as an indicator of a first-order EWPT, the
upper limit coming from previous lattice work.

Combining different values of $\mu$, we indicate the relative number
of points with a first-order phase transition as a heat map in
Fig.~\ref{fig:separate_heatmaps}, for three separate values of
$\tan(\beta)$. The majority of our points reside in the region
$m_{A_0} > m_{H_0} +m_Z$, in accordance with
Refs.~\cite{Dorsch:2014qja,Dorsch:2017nza} (see, however,
Refs.~\cite{Basler:2016obg,Basler:2017uxn}). In our framework,
sufficiently strong interactions with the second doublet are necessary
to bring $x$ down from its SM value of $x>0.11$.  Although the
relation between the 4D inputs and $x$ is complicated by the
diagonalization, a mass hierarchy between $H_0$ and $A_0$ generically
results in large portal couplings $\lambda_{3-5}$ and small values of
$x$ in the upper region. However, at small $\tan(\beta)$ we also see a
considerable number of points in regions where this does not hold.

\begin{figure}
\begin{center}
 \includegraphics[width=0.36\textwidth]{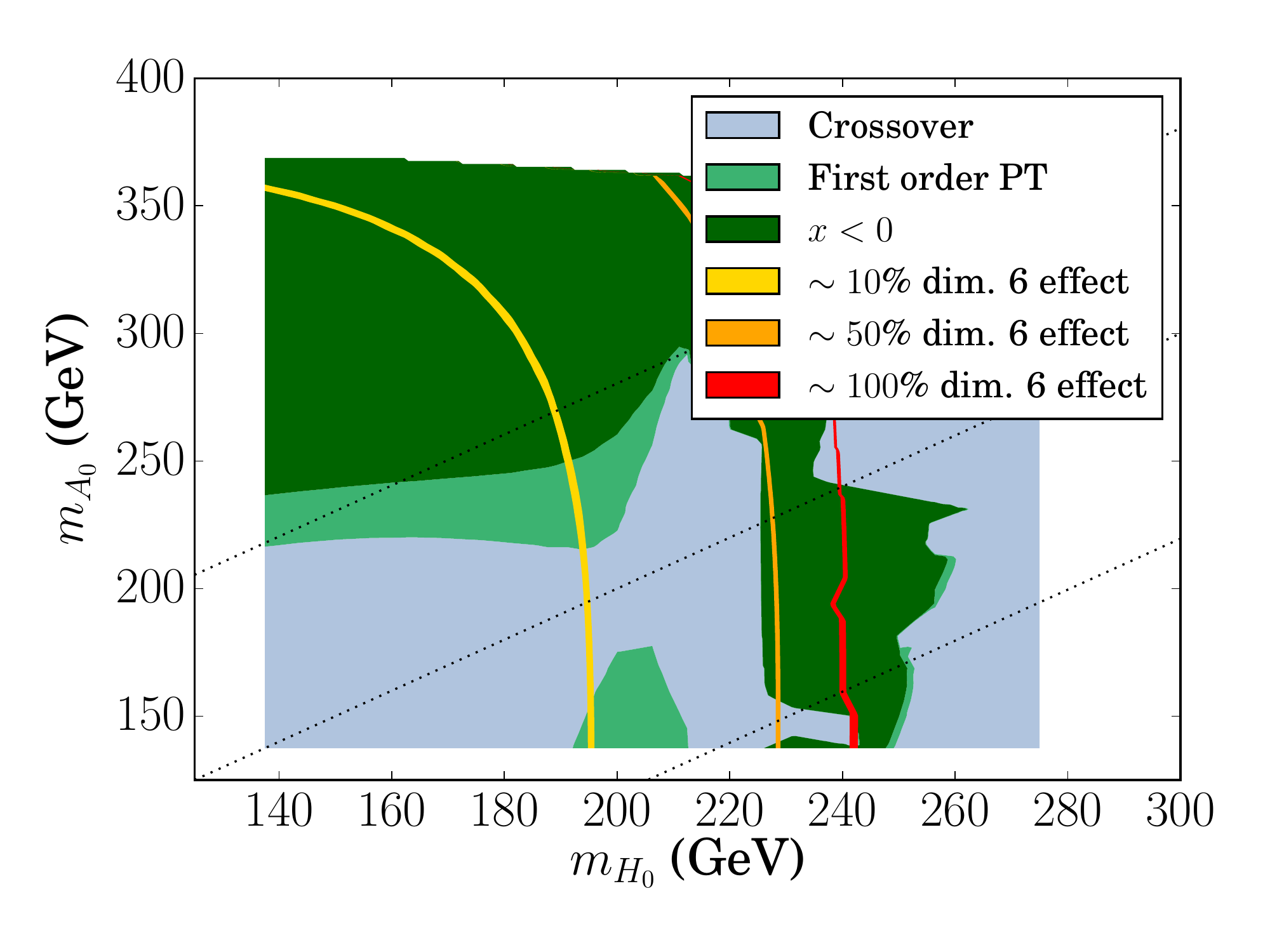} \\
 \includegraphics[width=0.36\textwidth]{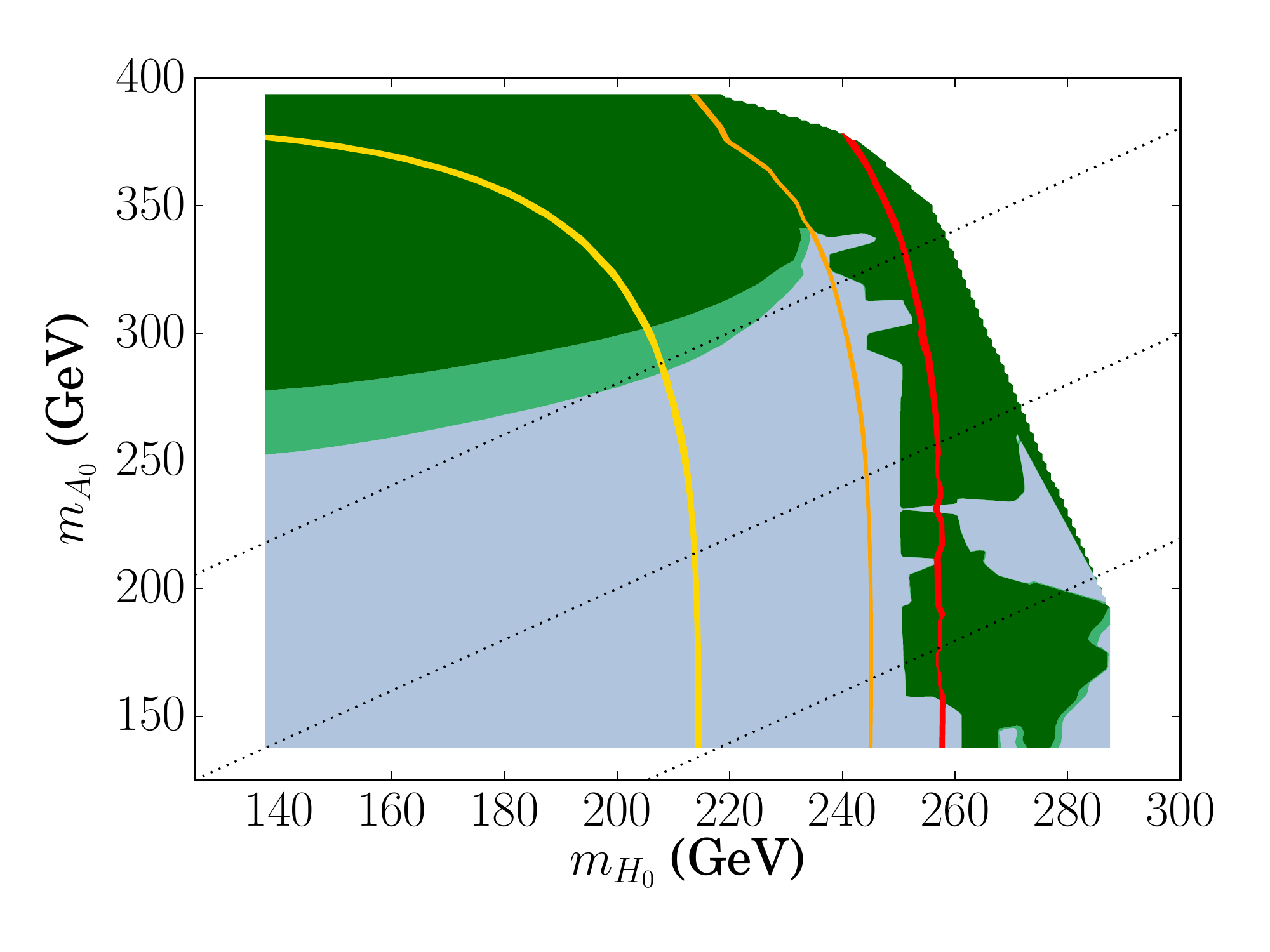}
\end{center}
\caption{\label{fig:slices} Slices with different values of $\mu$,
  $\mu= 50$ GeV (top) and 75 GeV (bottom), and fixed $\tan(\beta) =
  2.0$. The validity of DR is estimated by showing the relative effect
  of the neglected six-dimensional operators
      $\mathcal{O}^{(6)}_{A,1}$, $\mathcal{O}^{(6)}_{A,2}$. The
  white regions are either unphysical or there is no transition. At
  large $m_{H_0}$ the effects of six-dimensional operators render the
  first DR step unreliable.}
\end{figure}

In Fig.~\ref{fig:slices}, we show a breakdown of the heatmap plot with
fixed $\tan (\beta)=2.0$ for two values of $\mu$. We include here an
estimate of the effect of two of the neglected six-dimensional
  operators $\mathcal{O}^{(6)}_{A,1}$ and $\mathcal{O}^{(6)}_{A,2}$
produced when the superheavy scale is integrated out. Generally,
decreasing values of $x$ correspond to increasing importance of
six-dimensional terms: when the effect of these terms becomes large, DR
breaks down. These plots also show how the lower first-order region
disappears as $\mu$ increases.

We have found by explicit computation that the negative-$x$ region at
large $m_{A_0}$ is due to the omission of the six-dimensional operator
$\mathcal{O}^{(6)}_B$ in the last DR step that, although
inversely proportional to the heavy doublet mass, obtains sizable
contributions from the large couplings. We estimate its effect by
computing the dominant tree-level diagram contributing to the operator
coefficient (see \cite{Laine:1996ms} and the Supplemental Material)
and determining the two-loop effective potential in the final
effective theory with this operator included
(cf.~Refs.~\cite{Farakos:1994kx,Kajantie:1995dw}).  We stress,
however, that the effective potential is only a tool for estimating
errors from omitted six-dimensional operators; our results concerning
the phase transition are obtained using the non-perturbative phase
diagram of \cite{Kajantie:1995kf,Kajantie:1996qd}.

\begin{figure}
\begin{center}
\includegraphics[width=0.45\textwidth]{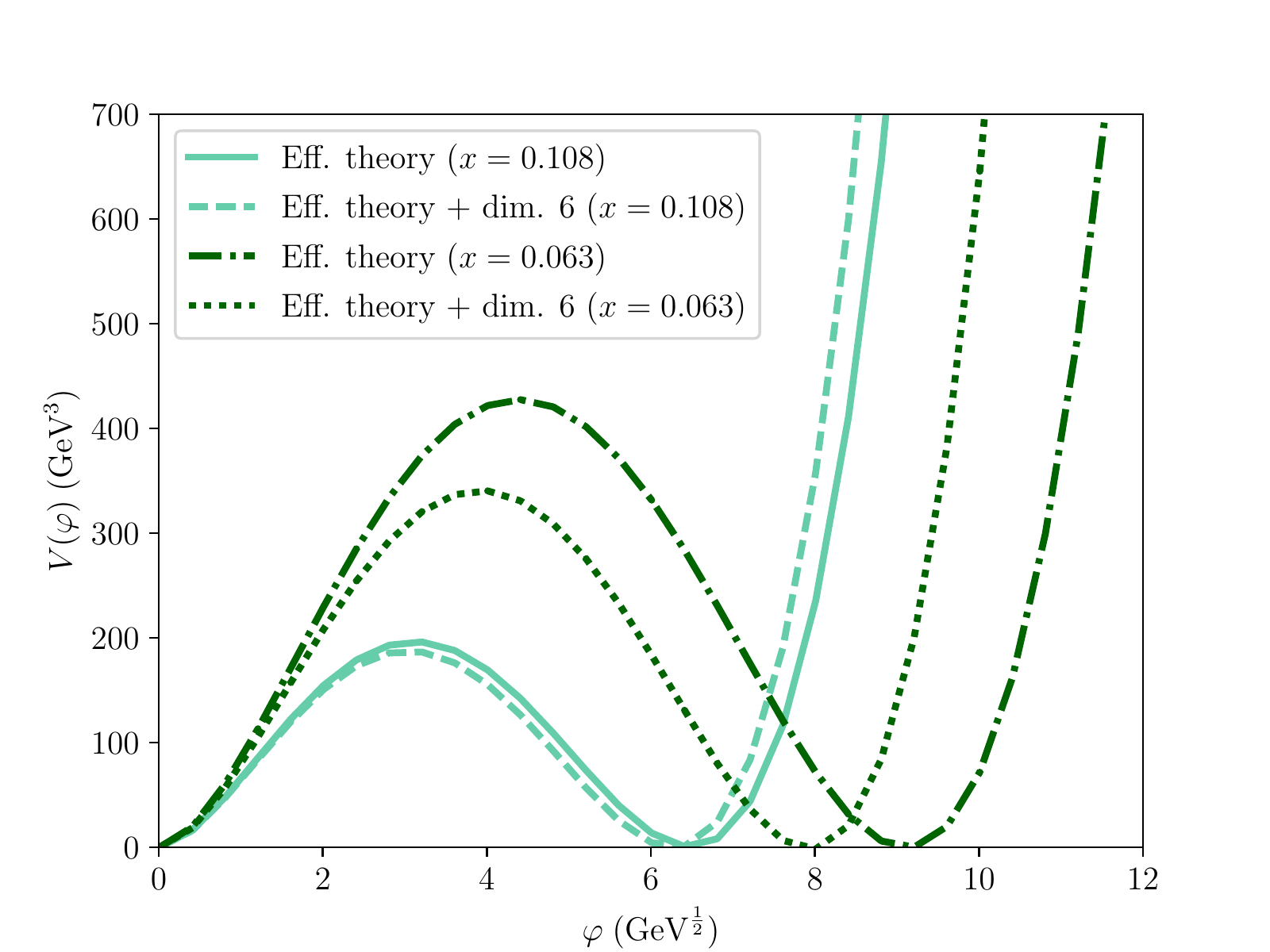}
\end{center}
\caption{\label{fig:Veff3d} Two-loop effective potential in the final
  effective theory with the dominant six-dimensional operator
  $\mathcal{O}^{(6)}_B$ of the last DR step included, evaluated
  at the critical temperature. At small $x$, integrating out the
  second doublet causes significant error, as is seen from the shift
  in the potential minimum. }
\end{figure}

In Fig.~\ref{fig:Veff3d}, the effective potential is depicted at two
values\footnote{The exact input parameters used were $m_{A_0} = 270\, \mathrm{GeV}$ ($x = 0.108$) and $m_{A_0} = 280\, \mathrm{GeV}$ ($x = 0.063$), with $\tan(\beta) = 2,\ m_{H_0}=180\, \mathrm{GeV}$, $\mu=75\, \mathrm{GeV}$, $m_{H^\pm} = m_{A_0}$ and $\cos(\beta-\alpha) = 0$ for both cases.} of $x$, both with and without the effects of the six-dimensional
operator $\mathcal{O}^{(6)}_B$. The field $\varphi$ is the 3D
background field, defined via $\langle \phi \rangle_{\td} =
\frac{\varphi}{\sqrt{2}}
\begin{pmatrix} 0 & 1 \end{pmatrix}^\mathrm{T}$ and related to 4D fields as described in the Supplemental
Material. The figure demonstrates how at $x=0.108$---near the
crossover boundary---the six-dimensional operator
$\mathcal{O}^{(6)}_B$ has a negligible impact on the potential,
while for $x=0.068$ (which corresponds to
$\phi_\mathrm{c}/T_\mathrm{c} \approx 0.7$) the effect is already
sizable, continuing to grow as $x$ decreases. Hence integrating out
the heavier doublet is expected to be a valid approximation when the
transition is of weakly first order, but becomes increasingly
challenged near the strong transition limit of $x \lesssim
0.04$. While we expect our results to be qualitatively robust even
there, reaching quantitatively accurate results for very small $x$
clearly calls for simulations with two dynamical doublets, which we
leave for future work.

Experimental constraints on the 2HDM parameter space depend strongly
on the way in which fermions couple to the Higgs doublets. With the
exception of the top quark, other Yukawa couplings have little effect
on our EWPT analysis, and we have still to indicate whether we are considering
Type I (all quarks couple to $\phi_2$) or Type II (up-type quarks
couple to $\phi_2$, down-type to $\phi_1$) 2HDM.  The most stringent
constraints come from flavour physics, where $B$-decays set the bound
$m_{H^\pm} \gtrsim 580$ GeV for the charged Higgs mass in Type
II~\cite{Misiak:2017bgg}. Assuming that $m_H^\pm$ is degenerate with
$m_{A_0}$ in accordance with EW precision tests, this rules out our
regions of first-order EWPT in Type II, but no such lower bound exists
in Type I for $\tan\beta \geq
2$~\cite{Mahmoudi:2009zx,Misiak:2017bgg}.

Additional restrictions come from direct searches for neutral Higgses
at the LHC~\cite{Aad:2014vgg}. For Type I, the $H_0~\rightarrow~\tau\tau$ cross section is suppressed by $\cot^2 \beta$, and our
choices of $\tan\beta$ are within current experimental
bounds. Finally, we have verified that the mass range we scan in is
allowed by measurements of the $h \rightarrow \gamma\gamma$
decay~\cite{CMS:2015kwa}, as well as the relatively recent search for
$A_0 \rightarrow Z h$ processes~\cite{Aad:2015wra}. Having not scanned
in the hidden-Higgs region where constraints from charged-scalar
searches become important~\cite{Pierce:2007ut}, we conclude that our
first-order EWPT regions are currently not ruled out by experiments if
a Type I 2HDM is assumed.

\section{Discussion}

It is a shortcoming of present-day particle
cosmology that it is still impossible to reliably determine the nature
and strength of the EWPT for a given BSM scenario. This information
would be valuable not only for EWBG, but also for gravitational wave
physics, as a first-order EWPT would leave an imprint in the
sensitivity range of the LISA mission and other proposed
gravitational-wave detectors~\cite{Caprini:2015zlo}.

We have taken a step towards remedying the situation by studying the
mapping of the phase diagram of one viable BSM theory, the 2HDM. Our
results concern the EWPT in the alignment limit $\cos(\beta-\alpha)$ =
0. Our work so far supports the idea that the primary signature of a
first order transition in this theory is indeed $m_{A_0} > m_{H_0} +
m_Z$, as suggested by Refs.~\cite{Dorsch:2014qja,Dorsch:2017nza}.

The techniques discussed in this paper can be applied, with suitable
modifications, to a host of other models where a substantial region of
parameter space can be mapped onto the three-dimensional theory of the
minimal Standard Model. In the future, our aim is to perform a
thorough comparison of perturbative and non-perturbative results in
the 2HDM by keeping both doublets dynamical in the effective
theory. Similar projects to study the EWPT and benchmark the accuracy
of perturbation theory are already underway in the Standard Model
augmented by a real singlet~\cite{singlet_scans} or triplet
field~\cite{Niemi:2018asa}; the EWPT has been perturbatively analysed
for the former in Refs.~\cite{Beniwal:2017eik,Chen:2017qcz}, and for
the latter in Ref.~\cite{Patel:2012pi}.

\begin{acknowledgments}
  The authors would like to thank Tom\'a\v{s} Brauner, Mark
Hindmarsh, Stephan J. Huber, Kimmo Kainulainen, Keijo Kajantie, Venus
Keus, Mikko Laine, Jose M. No and Kari Rummukainen for discussions. TT has been supported by the Vilho, Yrj\"{o}
and Kalle V\"{a}is\"{a}l\"{a} Foundation. TG, LN, TT, and AV have been
supported by the Academy of Finland grant no.~1303622, as well as by
the European Research Council grant no.~725369. LN was also supported
by the Academy of Finland grant no.~308791. DJW (ORCID ID
0000-0001-6986-0517) was supported by Academy of Finland grant
no.~286769.
\end{acknowledgments}

%%%%%%%%%%%%%%%%%%%%%%%%%%%%%%%%%%%%%%%%%%%%%%%%%%%%%%

%

\begin{widetext}

\newpage
\allowdisplaybreaks
\appendix
\begin{center}{\large \textbf{Supplemental Material}} \end{center}

\section{Dimensional reduction of 2HDM}

In this Supplemental Material we collect the matching relations
between the full four-dimensional theory and effective theories. A
detailed derivation can be found in Ref.~\cite{Gorda:2018hvi}.

\subsection{Three-dimensional effective theories}

We denote the fields of the effective theories with the same symbols
as those of the four-dimensional theory, even though their
normalisation is different and will affect the mapping between full
and effective theories. These normalisations between 4D and 3D
fields have been listed below.

The schematic form of classical Lagrangian density of the effective
theory was given in Eq.~(\ref{eq:full_3d_lag}) of the main paper. The
temporal part reads
\begin{multline}
\mathcal{L}^{(3)}_\text{temporal}
=\frac12(D_rA^a_0)^2+\frac12m_D^2A^a_0A^a_0+\frac12(\partial_rB_0)^2+\frac12m_D'^2B_0^2+\frac14\kappa_1(A^a_0A^a_0)^2+\frac14\kappa_2 B_0^4  \\
+\frac14\kappa_3 A^a_0A^a_0B_0^2+h_{1}\he\phi_1\phi_1 A^a_0A^a_0+h_{2}\he\phi_1\phi_1 B_0^2+h_{3}B_0\he\phi_1\vec A_0\cdot\vec\tau\phi_1 + h_{4}\he\phi_2\phi_2 A^a_0A^a_0 \\
+h_{5}\he\phi_2\phi_2 B_0^2+h_{6} B_0\he\phi_2\vec A_0\cdot\vec\tau\phi_2 
+ \frac12(\partial_rC^\alpha_0)^2+\frac12m_D''^2C^\alpha_0C^\alpha_0+ \omega_3 C^\alpha_0C^\alpha_0 \he\phi_2\phi_2.
\end{multline}
Here the covariant derivative of an isospin triplet reads $D_rA^a_0 =
\partial_r A^a_0 + g_3 \epsilon^{abc}A^b_rA^c_0$, and for the temporal
gluon $C^\alpha_0$ ordinary derivative is used instead of covariant
derivative as gluons are discarded for only contributing at a higher
order~\cite{Brauner:2016fla}.

After the heavy temporal scalars have been integrated out, their
effects are encoded by the parameters and fields of a new theory where
the parameters are denoted with a bar as $\bar{g}_3, {\bar{g}'}_3,
\bar{\mu}^2_{11,3}$ etc. In this theory, the phase transition takes
place close to a point where the mass matrix has zero eigenvalue, and
then generically in the diagonal basis the other mass parameter is
heavy. By performing a unitary transformation, one can diagonalise the
scalar potential. Denoting $\Omega \equiv \; \sqrt{(\bar{\mu}^2_{11,3}
  - \bar{\mu}^2_{22,3})^2 + 4 \bar{\mu}^{2*}_{12,3}
  \bar{\mu}^{2}_{12,3}}$, this transformation reads\footnote{Assuming $\RE \bar{\mu}^2_{12,3} > 0$;
  otherwise, $\alpha$ and
  $\delta$ change sign.}
\begin{equation}
\begin{pmatrix}
    \phi_1 \\
    \phi_2
  \end{pmatrix}
\equiv
\begin{pmatrix}
    \alpha & \beta \\
    \gamma & \delta
\end{pmatrix}
\begin{pmatrix}
    \theta \\
    \phi
  \end{pmatrix},
\end{equation}
where
\begin{align}
\alpha & \equiv \frac{2}{\sqrt{4 + \left|\frac{(\bar{\mu}^2_{22,3}
      - \bar{\mu}^2_{11,3}) + \Omega}{ \bar{\mu}^{2*}_{12,3}
    }\right|^2}},
&
\beta & \equiv \; \frac{(\bar{\mu}^2_{11,3} - \bar{\mu}^2_{22,3}
  - \Omega^*)}{\bar{\mu}^2_{12,3} \sqrt{4 +
    \left|\frac{(\bar{\mu}^2_{22,3} - \bar{\mu}^2_{11,3}) + \Omega}{
      \bar{\mu}^{2*}_{12,3} }\right|^2}}  \nonumber \\
\gamma & \equiv \frac{2}{\sqrt{4 + \left|\frac{-(\bar{\mu}^2_{22,3}
      - \bar{\mu}^2_{11,3}) + \Omega}{ \bar{\mu}^{2*}_{12,3}
    }\right|^2}},
&
\delta & \equiv \; \frac{(\bar{\mu}^2_{11,3} -
  \bar{\mu}^2_{22,3} + \Omega^*)}{\bar{\mu}^2_{12,3} \sqrt{4 +
    \left|\frac{-(\bar{\mu}^2_{22,3} - \bar{\mu}^2_{11,3}) + \Omega}{
      \bar{\mu}^{2*}_{12,3} }\right|^2}}.
\end{align}
The mass parameters in the diagonal basis read
\begin{align}
\widetilde{\mu}^2_\phi & = \frac{1}{2}( \bar{\mu}^2_{11,3} +
\bar{\mu}^2_{22,3} - \Omega),
&
\widetilde{\mu}^2_\theta & = \frac{1}{2}( \bar{\mu}^2_{11,3} + \bar{\mu}^2_{22,3} + \Omega).
\end{align}
Generally $\widetilde{\mu}^2_\theta$ is heavy when
$\widetilde{\mu}^2_\phi$ is light, and therefore the field $\theta$
can be integrated out. The scalar self-couplings in the diagonal basis
are given by
\begin{align}
  \begin{pmatrix}
    \widetilde{\lambda}_1 \\
    \widetilde{\lambda}_2 \\
    \widetilde{\lambda}_3 \\
    \widetilde{\lambda}_4 \\
    \widetilde{\lambda}_5 /2 \\
    \widetilde{\lambda}_6 \\
    \widetilde{\lambda}_7
  \end{pmatrix}
=
    M \cdot 
\begin{pmatrix}
    \bar{\lambda}_{1,3} \\
    \bar{\lambda}_{2,3} \\
    \bar{\lambda}_{3,3} \\
    \bar{\lambda}_{4,3} \\
    \bar{\lambda}_{5,3}/2 \\
    \bar{\lambda}^*_{5,3}/2 %\\
  \end{pmatrix},
\end{align}
where
\begin{align}
M \equiv 
\begin{pmatrix}
     |\beta|^4 & |\delta|^4 & |\beta|^2|\delta|^2 & |\beta|^2|\delta|^2 & (\beta^*\delta)^2 & (\beta \delta^*)^2 \\
    |\alpha|^4 & |\gamma|^4 & |\alpha|^2|\gamma|^2 &
    |\alpha|^2|\gamma|^2  & (\alpha^*\gamma)^2 &
    (\alpha\gamma^*)^2 \\
    2 |\alpha|^2|\beta|^2 & 2 |\gamma|^2|\delta|^2 & |\alpha|^2|\delta|^2 + |\beta|^2|\gamma|^2  &2\RE( \alpha \beta^* \gamma^* \delta)  & 2 \alpha^* \beta^* \gamma \delta & 2 \alpha\beta\gamma^*\delta^*  \\
    2|\alpha|^2|\beta|^2 & 2|\gamma|^2|\delta|^2 & 2\RE(\alpha\beta^*\gamma^*\delta)  & |\alpha|^2 |\delta|^2 + |\beta|^2 |\gamma|^2   & 2 \alpha^*\beta^*\gamma\delta & 2 \alpha\beta\gamma^*\delta^*  \\
    (\alpha\beta^*)^2 & (\gamma\delta^*)^2 & \alpha\beta^*\gamma\delta^* & \alpha\beta^*\gamma\delta^* & (\beta^*\gamma)^2 &  (\alpha\delta^*)^2 \\
    2 |\beta|^2\alpha\beta^* & 2 |\delta|^2 \gamma\delta^* & \beta^*\delta^*(\beta\gamma+\alpha\delta) &\beta^*\delta^*(\beta\gamma+\alpha\delta) & 2 \beta^*\gamma\beta^*\delta & 2 \alpha\beta\delta^*\delta^* \\
    2 |\alpha|^2\alpha^*\beta & 2 |\gamma|^2 \gamma^* \delta & \alpha^*\gamma^*(\beta\gamma+\alpha\delta) &  \alpha^*\gamma^*(\beta\gamma+\alpha\delta) & 2 \alpha^* \gamma \alpha^*\delta & 2 \alpha\beta\gamma^*\gamma^* \\
  \end{pmatrix}.
\end{align}

The scalar potential in the diagonal basis reads
\begin{multline}
V(\phi,\theta) = \widetilde{\mu}^2_{\phi} \he\phi \phi +
\widetilde{\mu}^2_{\theta} \he\theta \theta + \widetilde{\lambda}_1
(\he\phi \phi)^2 + \widetilde{\lambda}_2 (\he\theta \theta)^2 +
\widetilde{\lambda}_3 (\he\phi \phi)(\he\theta \theta) +
\widetilde{\lambda}_4 (\he\phi \theta)(\he\theta \phi) \\ +
\frac{\widetilde{\lambda}_5}{2} (\he\phi \theta)^2 +
\frac{\widetilde{\lambda}^*_5}{2} (\he\theta \phi)^2 +
\widetilde{\lambda}_6 (\he\phi \phi)(\he\phi \theta) +
\widetilde{\lambda}^*_6 (\he\phi \phi)(\he\theta \phi) +
\widetilde{\lambda}_7 (\he\theta \theta)(\he\theta \phi) +
\widetilde{\lambda}^*_7 (\he\theta \theta)(\he\phi \theta),
\end{multline}
where $\phi$ and $\theta$ are light and heavy fields, respectively.

When the heavy doublet $\theta$ has been integrated out, the final
effective theory is same as in that of the SM, as given in
Eq.~(\ref{eq:3D}) of the main paper.

\subsection{Matching relations and normalisations of fields}

Our calculations are carried out in the $\msbar$ scheme. We use the
following notation:
\begin{align}
N_d & =2, \nonumber \\
L_b & \equiv 2 \ln \left( \frac{\Lambda}{T} \right) -2 [\ln(4\pi)-\gamma], \\ 
L_f & \equiv L_b + 4 \ln 2, \nonumber \\
c & \equiv \frac{1}{2} \left( \ln \left( \frac{8\pi}{9} \right) + \frac{\zeta'(2)}{\zeta(2)} - 2 \gamma \right), \nonumber
\end{align}
where $\Lambda$ is the renormalisation scale of the 4D theory and
$\gamma$ is the Euler-Mascheroni constant.

The normalisations relating three- and four-dimensional fields read
\begin{align}
A_{\td,0}^2 & = \frac{A_{\fd,0}^2}{T} \left\{1+\frac{g^2}{(4\pi)^2}\left[\frac{N_d-26}{6}L_b+\frac{1}{3}(8+N_d)+\frac{4N_f}{3}(L_f-1)\right]\right\},  \nonumber \\
A_{\td,r}^2 & = \frac{A_{\fd,r}^2}{T} \left[1+\frac{g^2}{(4\pi)^2}\left(\frac{N_d-26}{6}L_b-\frac{2}{3}+\frac{4N_f}{3}L_f\right)\right], \nonumber \\
B_{\td,0}^2 & = \frac{B_{\fd,0}^2}{T} \left\{1+\frac{g'^2}{(4\pi)^2}\left[N_d\left(\frac{L_b}{6}+\frac{1}{3}\right)+\frac{20N_f}{9}(L_f-1)\right]
\right\}, \nonumber \\
B_{\td,r}^2 & =\frac{B_{\fd,r}^2}{T} \left[1+
  \frac{g'^2}{(4\pi)^2}\left(N_d\frac{L_b}{6}+\frac{20N_f}{9}L_f\right)\right], \\
\left(\phi^{\dagger}_1\phi_1\right)_\td & =
\frac{\left(\phi^{\dagger}_1\phi_1\right)_\fd}{T} \left[1-\frac{1}{(4\pi)^2}\left(\frac{3}{4}(3g^2 + {g'}^2)L_b  \right)\right], \nonumber \\
\left(\phi^{\dagger}_2\phi_2\right)_\td & =
\frac{\left(\phi^{\dagger}_2\phi_2\right)_\fd}{T} \left[1-\frac{1}{(4\pi)^2}\left(\frac{3}{4}(3g^2 + {g'}^2)L_b - 3 g^2_{Y} L_f
\right)\right], \nonumber \\
\left(\phi^{\dagger}_1\phi_2\right)_\td & =
\frac{\left(\phi^{\dagger}_1\phi_2\right)_\fd}{T} \left[1-\frac{1}{(4\pi)^2}\left(\frac{3}{4}(3g^2 + {g'}^2)L_b - \frac{3}{2} g^2_{Y} L_f \right)\right]. \nonumber
\end{align}
For dimensional reduction, the required ingredients include matching
relations between the 4D and 3D theories, one-loop $\beta$ functions
(to make the matching relations renormalisation scale independent) and
finally the relations between $\overline{\rm MS}$-parameters and
physical quantities.

We use the tree-level relations, despite the fact that for consistent
$O(g^4)$ accuracy one should use the one-loop corrected
relations. This would require performing one-loop vacuum
renormalisation of the physical quantities. This is a non-trivial
task, and is left for the future. In the special case of the inert
doublet model, the one-loop vacuum renormalisation can be found in
Ref.~\cite{Laine:2017hdk}. Below we list all needed matching
relations, while $\beta$ functions and relations of
$\msbar$-parameters and physical quantities can be found in
Ref.~\cite{Gorda:2018hvi} with detailed derivations and explicit,
step-by-step intermediate results.

\subsubsection{Integration over superheavy scale}

A full $O(g^4)$-accurate dimensional reduction requires the evaluation
of the mass parameters at two-loop and couplings at one-loop
order. The results are listed below.
\begin{align}
m_D^2={}&g^2T^2\bigg(\frac{4+N_d}{6}+\frac{N_f}{3}\bigg),\\
m'^2_D={}&g'^2T^2\bigg(\frac{N_d}{6}+\frac{5N_f}{9}\bigg),\\
m''^2_D={}&g^2_s T^2\bigg(1+\frac{N_f}{6}\bigg),\\
g_3^2={}&g^2(\Lambda)T\bigg[1 +\frac{g^2}{(4\pi)^2}\bigg(\frac{44-N_d}{6}L_b+\frac{2}{3}-\frac{4N_f}{3}L_f\bigg)\bigg],\\
g'^2_3={}&g'^2(\Lambda)T\bigg[1 -\frac{g'^2}{(4\pi)^2}\bigg(\frac{N_d}{6}L_b+\frac{20N_f}{9}L_f\bigg)\bigg],\\
\kappa_1={}&T\frac{g^4}{16 \pi^2} \frac{16+N_d-4N_f}{3},\\
\kappa_2={}&T\frac{g'^4}{16\pi^2} \bigg(\frac{N_d}{3}-\frac{380}{81} N_f\bigg),\\
\kappa_3={}&T\frac{g^2g'^2}{16\pi^2}\bigg(2N_d-\frac{8}{3}N_f\bigg),\\
\notag
%%%%%%%
h_{1}={}&\frac{g^2(\Lambda)T}{4}\bigg(1+\frac{1}{(4\pi)^2}\bigg\{\bigg[\frac{44-N_d}{6}L_b+\frac{53}{6}-\frac{N_d}{3}-\frac{4N_f}{3}(L_f-1)\bigg]g^2+\frac{g'^2}{2} \\
& \quad \quad \quad \quad \quad \quad  \quad \quad \quad + 12\lambda_1  + 2 (2\lambda_3+\lambda_4) \bigg\} \bigg), \\
h_{2}={}&\frac{g'^2(\Lambda)T}{4}\bigg(1 +\frac{1}{(4\pi)^2}\bigg\{\frac{3g^2}{2}+\bigg[\frac{1}{2}-\frac{N_d}{6}\Big(2+L_b \Big)  -\frac{20N_f}{9}(L_f-1)\bigg]g'^2 \notag \\
& \quad \quad \quad \quad \quad \quad  \quad \quad \quad + 12\lambda_1  + 2 (2\lambda_3+\lambda_4) \bigg\} \bigg), \\
h_{3}={}&\frac{g(\Lambda)g'(\Lambda)T}{2}\bigg\{1+\frac{1}{(4\pi)^2}\bigg[-\frac{5+N_d}{6} g^2+ \frac{3-N_d}{6}g'^2+L_b\bigg(\frac{44-N_d}{12}g^2 -\frac{N_d}{12}g'^2\bigg)\notag \\
&-N_f(L_f-1)\bigg(\frac{2}{3}g^2+\frac{10}{9}g'^2\bigg) + 4 \lambda_1 + 2 \lambda_4 \bigg]\bigg\},\\
%%%%%%%%%%%%%%%%
h_{4}={}&\frac{g^2(\Lambda)T}{4}\bigg(1+\frac{1}{(4\pi)^2}\bigg\{\bigg[\frac{44-N_d}{6}L_b+\frac{53}{6}-\frac{N_d}{3}-\frac{4N_f}{3}(L_f-1)\bigg]g^2+\frac{g'^2}{2} -6 g_{Y}^2\\
& \quad \quad \quad \quad \quad \quad  \quad \quad \quad + 12\lambda_2  + 2 (2\lambda_3+\lambda_4) \bigg\} \bigg), \\
h_{5}={}&\frac{g'^2(\Lambda)T}{4}\bigg(1 +\frac{1}{(4\pi)^2}\bigg\{\frac{3g^2}{2}+\bigg[\frac{1}{2}-\frac{N_d}{6}\Big(2+L_b \Big)  -\frac{20N_f}{9}(L_f-1)\bigg]g'^2 - \frac{34}{3} g_{Y}^2 \notag \\
& \quad \quad \quad \quad \quad \quad  \quad \quad \quad + 12\lambda_2  + 2 (2\lambda_3+\lambda_4) \bigg\} \bigg), \\
h_{6}={}&\frac{g(\Lambda)g'(\Lambda)T}{2}\bigg\{1+\frac{1}{(4\pi)^2}\bigg[-\frac{5+N_d}{6} g^2+ \frac{3-N_d}{6}g'^2+L_b\bigg(\frac{44-N_d}{12}g^2 -\frac{N_d}{12}g'^2\bigg)\notag \\
&-N_f(L_f-1)\bigg(\frac{2}{3}g^2+\frac{10}{9}g'^2\bigg) + 2 g_{Y}^2 + 4 \lambda_2 + 2 \lambda_4 \bigg]\bigg\},\\
\omega_3={}&-\frac{2 T}{16 \pi^2} g^2_s g^2_{Y}, \\
\lambda_{1,3}={}&T\Bigg\{\lambda_1(\Lambda) + \frac{1}{(4\pi)^2}\bigg[\frac{1}{8}\Big(3g^4 + {g'}^4 +2 g^2{g'}^2 \Big)  \nonumber  \\
& -L_b \bigg(\frac{3}{16}\Big(3g^4 + {g'}^4 + 2 g^2{g'}^2 \Big) + \lambda^2_3 + \lambda_3 \lambda_4 + \frac{1}{2}\lambda^2_4 + \frac{1}{2}|\lambda_5|^2 - \frac{3}{2}\Big(3g^2+{g'}^2 -8 \lambda_1 \Big) \lambda_1 \bigg) \bigg]\Bigg\}, \\
\lambda_{2,3}={}&T\Bigg\{\lambda_2(\Lambda) + \frac{1}{(4\pi)^2}\bigg[\frac{1}{8}\Big(3g^4 + {g'}^4 +2 g^2{g'}^2 \Big) + 3 L_f \Big(g^4_{Y} - 2\lambda_2 g^2_{Y} \Big)\nonumber  \\
& -L_b \bigg(\frac{3}{16}\Big(3g^4 + {g'}^4 + 2 g^2{g'}^2 \Big) + \lambda^2_3 + \lambda_3 \lambda_4 + \frac{1}{2}\lambda^2_4 + \frac{1}{2}|\lambda_5|^2 - \frac{3}{2}\Big(3g^2+{g'}^2 -8 \lambda_2 \Big) \lambda_2 \bigg) \bigg]\Bigg\}, \\
\lambda_{3,3}={}&T\Bigg\{\lambda_3(\Lambda) + \frac{1}{(4\pi)^2}\bigg[\frac{1}{4}\Big(3g^4 + {g'}^4 -2 g^2{g'}^2 \Big) - 3 L_f \lambda_3  g^2_{Y} \nonumber  \\
& -L_b \bigg(\frac{3}{8}\Big(3g^4 + {g'}^4 - 2 g^2{g'}^2 \Big) + 2(\lambda_1 + \lambda_2)(3\lambda_3 + \lambda_4) + 2\lambda^2_3 +  \lambda^2_4 + |\lambda_5|^2 - \frac{3}{2}\Big(3g^2+{g'}^2\Big) \lambda_3 \bigg) \bigg]\Bigg\}, \\
\lambda_{4,3}={}&T\Bigg\{\lambda_4(\Lambda) + \frac{1}{(4\pi)^2}\bigg[ g^2{g'}^2  - 3 L_f \lambda_4  g^2_{Y}  \nonumber  \\
& -L_b \bigg(\frac{3}{2} g^2{g'}^2 + 2(\lambda_1 + \lambda_2)\lambda_4 + 2\lambda^2_4 + 4\lambda_3\lambda_4 + 4|\lambda_5|^2  - \frac{3}{2}\Big(3g^2+{g'}^2\Big) \lambda_4 \bigg) \bigg]\Bigg\} \\
\text{and} \quad \lambda_{5,3}={}&T\Bigg\{\lambda_5(\Lambda) + \frac{1}{(4\pi)^2}\bigg[  -3 L_f \lambda_5 g^2_{Y} -L_b \bigg( 2(\lambda_1 + \lambda_2 + 2 \lambda_3 + 3 \lambda_4)\lambda_5 - \frac{3}{2}\Big(3g^2+{g'}^2\Big) \lambda_5 \bigg) \bigg]\Bigg\}.
\end{align}
These relations
have been calculated previously in Ref.~\cite{Losada:1996ju}, with the
restriction of $\lambda_5$ being real rather than complex. We have
corrected two minor errors in the expressions for $\lambda_{4,3}$ and
$\lambda_{5,3}$ for the terms involving $\lambda_5$.

In the Standard Model, the $O(g^4)$ result for the 3D scalar mass
parameter reads:
\begin{align}
\Big(\mu^2_{22,3}\Big)_\text{SM} =& \mu^2_{22}(\Lambda) +\frac{T^2}{16}\Big(3g^2(\Lambda) + {g'}^2(\Lambda) + 4 g^2_Y(\Lambda) + 8 \lambda_2(\Lambda) \Big) \nonumber \\
& + \frac{1}{16\pi^2} \bigg\{ \mu^2_{22}\bigg( \Big(\frac{3}{4}(3g^2 + {g'}^2) - 6 \lambda_2 \Big)L_b - 3 g^2_Y L_f \bigg) \nonumber \\
& + T^2 \bigg( \frac{167}{96}g^4 + \frac{1}{288}{g'}^4 - \frac{3}{16}g^2{g'}^2 + \frac{1}{4}\lambda_2(3g^2+{g'}^2) \nonumber \\
& + L_b \Big( \frac{17}{16}g^4 - \frac{5}{48}{g'}^4 - \frac{3}{16}g^2{g'}^2 + \frac{3}{4}\lambda_2(3g^2+{g'}^2) - 6 \lambda^2_2 \Big) \nonumber \\
& + \frac{1}{T^2}\Big( c + \ln(\frac{3T}{\Lambda_{\td}}) \Big)\Big( \frac{39}{16}g^4_3 + 12g^2_3 h_{4} - 6 h^2_{4} + 9 g^2_3 \lambda_{2,3} - 12\lambda^2_{2,3} \nonumber \\
& -\frac{5}{16}{g'}^4_3 - \frac{9}{8}g^2_3 {g'}^2_3 - 2h^2_{5} - 3h^2_6 + 3 {g'}^2_3 \lambda_{2,3} \Big) \nonumber \\
& - g^2_Y \Big(\frac{3}{16}g^2 + \frac{11}{48}{g'}^2 + 2 g^2_s \Big) + (\frac{1}{12}g^4 + \frac{5}{108}{g'}^4)N_f \nonumber \\
& + L_f \Big( g^2_Y \Big(\frac{9}{16}g^2 + \frac{17}{48}{g'}^2 + 2 g^2_s - 3 \lambda_2 \Big) +\frac{3}{8}g^4_Y - (\frac{1}{4}g^4 + \frac{5}{36}{g'}^4 ) N_f \Big) \nonumber \\
& + \ln(2) \Big( g^2_Y \Big(-\frac{21}{8}g^2 - \frac{47}{72}{g'}^2 + \frac{8}{3} g^2_s + 9 \lambda_2 \Big) -\frac{3}{2}g^4_Y + (\frac{3}{2}g^4 + \frac{5}{6}{g'}^4 ) N_f \Big) \bigg) \bigg\}.
\end{align}
This result can be found also in Ref.~\cite{Kajantie:1995dw}, apart
from the two-loop contributions involving ${g'}$, as in that paper
these were assumed to scale as ${g'}\sim g^{\frac{3}{2}}$.

In the 2HDM, the 3D scalar mass parameters read:
\begin{align}
\Big(\mu^2_{22,3}\Big)_\text{2HDM} =& \Big(\mu^2_{22,3}\Big)_\text{SM} + \frac{T^2}{12}\Big(2\lambda_3(\Lambda)+\lambda_4(\Lambda) \Big) \nonumber \\
& + \frac{1}{16\pi^2}\bigg\{  \mu^2_{11}\Big( -L_b(2\lambda_3 + \lambda_4) \Big) \nonumber \\
& + T^2 \bigg( \frac{5}{48} g^4 + \frac{5}{144} {g'}^4 + \frac{1}{24}(3g^2 + {g'}^2)( 2\lambda_3 + \lambda_4) \nonumber \\
&+ \frac{1}{T^2}\Big(c + \ln\big(\frac{3T}{\Lambda_{\td}} \big) \Big)\Big( -\frac{1}{8}(3 g^4_3 + {g'}^4_3) + \frac{1}{2}(3g^2_3 + {g'}^2_3)(2\lambda_{3,3} + \lambda_{4,3}) \nonumber \\ 
& - 2 (\lambda^2_{3,3} + \lambda_{3,3} \lambda_{4,3} + \lambda^2_{4,3})  - 3 |\lambda_{5,3}|^2 \Big) \nonumber \\
&+ L_b \Big( -\frac{7}{32}g^4 - \frac{7}{96}{g'}^4
-\frac{1}{2}(\lambda_1+\lambda_2)(2\lambda_3+\lambda_4) \nonumber \\
& -\frac{5}{6}\lambda^2_3 - \frac{7}{12}\lambda^2_4 -\frac{5}{6} \lambda_3 \lambda_4 - \frac{3}{4}|\lambda_5|^2 +\frac{1}{8}(3g^2+{g'}^2) \big( 2\lambda_3 + \lambda_4 \big)  \Big) \nonumber \\
&+ \Big( - \frac{1}{4}g^2_Y \big(2\lambda_3 + \lambda_4 \big)  \Big)L_f  \bigg)  \nonumber \bigg\},
\end{align}
and
\begin{align}
\mu^2_{11,3} =& \mu^2_{11}(\Lambda) +\frac{T^2}{16}\Big(3g^2(\Lambda) + {g'}^2(\Lambda) + 8 \lambda_1(\Lambda) +\frac{4}{3}\Big(2\lambda_3(\Lambda) + \lambda_4(\Lambda) \Big)  \Big) \nonumber \\
& + \frac{1}{16\pi^2} \bigg\{L_b \bigg( \Big(\frac{3}{4}(3g^2 + {g'}^2) - 6 \lambda_1 \Big)\mu^2_{11} - (2\lambda_3 + \lambda_4)\mu^2_{22} \bigg) \nonumber \\
& + T^2 \bigg( \frac{59}{32}g^4 + \frac{11}{288}{g'}^4 - \frac{3}{16}g^2{g'}^2 + \frac{1}{4}\lambda_1(3g^2+{g'}^2) +\frac{1}{24}(3g^2 + {g'}^2)( 2\lambda_3 + \lambda_4) \nonumber \\
& + L_b \Big( \frac{27}{32}g^4 - \frac{17}{96}{g'}^4 - \frac{3}{16}g^2{g'}^2 + \frac{1}{8}(3g^2+{g'}^2)(6\lambda_1 + 2\lambda_3 + \lambda_4) -\frac{1}{2}(\lambda_1+\lambda_2)(2\lambda_3 + \lambda_4) \nonumber \\
& - 6 \lambda^2_1 - \frac{5}{6}\lambda^2_3 - \frac{5}{6}\lambda_3 \lambda_4 - \frac{7}{12}\lambda^2_4 - \frac{3}{4}|\lambda_5|^2  \Big) \nonumber \\
& + \frac{1}{T^2}\Big( c + \ln(\frac{3T}{\Lambda_{\td}}) \Big)\Big( \frac{33}{16}g^4_3 + 12g^2_3 h_{1} - 6 h^2_{1} + 9 g^2_3 \lambda_{1,3} - 12\lambda^2_{1,3} \nonumber \\
& -\frac{7}{16}{g'}^4_3 - \frac{9}{8}g^2_3 {g'}^2_3 - 2h^2_{2} - 3h^2_{3} + 3 {g'}^2_3 \lambda_{1,3} + \frac{1}{2}(3g^2_3 + {g'}^2_3)(2\lambda_{3,3} + \lambda_{4,3}) \nonumber \\
&- 2 (\lambda^2_{3,3} + \lambda_{3,3} \lambda_{4,3} + \lambda^2_{4,3}) - 3 |\lambda_{5,3}|^2 \Big) \nonumber \\
& + (\frac{1}{12}g^4 + \frac{5}{108}{g'}^4)N_f + L_f \Big( -\frac{1}{4}g^2_Y(2\lambda_3 + \lambda_4)  - (\frac{1}{4}g^4 + \frac{5}{36}{g'}^4 ) N_f \Big) \nonumber \\
& + \ln(2) \Big( \frac{3}{2} g^2_Y \Big(2\lambda_3 + \lambda_4 \Big) + (\frac{3}{2}g^4 + \frac{5}{6}{g'}^4 ) N_f \Big) \bigg) \bigg\},
\end{align}
and finally
\begin{equation}
\mu^2_{12,3} = \mu^2_{12}(\Lambda) + \frac{1}{16\pi^2}\bigg\{ L_b
\bigg(\Big(\frac{3}{4}(3g^2+{g'}^2) -\lambda_3 -2 \lambda_4 \Big)
\mu^2_{12} - 3 \lambda_5 \mu^{2*}_{12} \bigg) - \frac{3}{2}g^2_Y
\mu^2_{12} L_f \bigg\}.
\end{equation}

We emphasise that all matching relations in the first step of DR are
independent of the four-dimensional theory renormalisation scale
$\Lambda$ to the order we consider here, which can be seen by applying
the $\beta$ functions to the tree-level terms. This serves as a
crosscheck of the correctness of our calculation. In practice, we fix
$\Lambda=4\pi e^{-\gamma}T \approx 7T$, and have verified that the
choice of $\Lambda$ has negligible effect on our results.

These relations for the two-loop mass parameters have been calculated
previously in Ref.~\cite{Andersen:1998br}, under the assumption of
$\lambda_5$ being real (see also Ref.~\cite{Losada:1998at}).  However,
in Ref.~\cite{Andersen:1998br} there are again minor errors
propagating from the one-loop result of Ref.~\cite{Losada:1996ju}.

Note that in the coefficients of $ c +
\ln(\frac{3T}{\Lambda_{\td}})$, we have used a higher order result that
can be obtained by solving the running directly in the 3D theory;
because the 3D theory is super-renormalisable, this is the exact
dependence of the 3D theory renormalisation group scale $\Lambda_{\td}$. The mass
counterterms in the 3D theory can be found in
Ref.~\cite{Gorda:2018hvi}. We have fixed the renormalisation
scale of the 3D theory as $\Lambda_{\td} = {g}^2_3$, but it is possible
that some other choice would be more suitable for resumming higher-order
corrections, especially when the couplings are allowed to be large. We 
leave a quantitative analysis of renormalisation group effects in 3D 
for future work.

We estimate the validity of the effective theory constructed using 
the above matching relations by evaluating to
order $O(g^6)$ the terms $\mathcal{O}^{(6)}_{A,1} = \Lambda_{A,1} (\phi^\dagger_1
\phi_1)^3_{\td}$ and $\mathcal{O}^{(6)}_{A,2} = \Lambda_{A,2} (\phi^\dagger_2
\phi_2)^3_{\td}$, where the coefficients are given by
\begin{equation}
  \label{eq:dim6op1}
\Lambda_{A,1} = \frac{\zeta(3)}{3 (4\pi)^4}\Big( 30 \lambda^3_1 + \frac{1}{4} \lambda^3_3 + \lambda^3_+ + \lambda^3_- + \frac{3}{32}g^6 + \frac{3}{64}(g^2+{g'}^2)^3 \Big),
\end{equation}
\begin{equation}
  \label{eq:dim6op2}
\Lambda_{A,2} = \frac{\zeta(3)}{3 (4\pi)^4}\Big( 30 \lambda^3_2 + \frac{1}{4} \lambda^3_3 + \lambda^3_+ + \lambda^3_- + \frac{3}{32}g^6 + \frac{3}{64}(g^2+{g'}^2)^3 - \frac{21}{2} g^6_Y \Big),
\end{equation}
where $\lambda_\pm \equiv \frac{1}{2}(\lambda_3 + \lambda_4 \pm
\lambda_5)$. The last term in the above equation is the contribution
from the top quark. Comparing the magnitude of the other terms to that
term yields a rough estimate of the importance of neglected
six-dimensional operators.

\subsubsection{Integration over heavy scale}

Here we list the matching results for the parameters in the simplified
3D effective theories, where the heavy temporal scalars and the heavy
second doublet have been integrated out. The two-loop contributions to
mass parameters are highlighted with subscripts. One could give the
coefficients of logarithmic contributions in terms of the running in
the final theory here as well, but we have omitted this for
simplicity.
\begin{align}
\bar{g}^2_3 = & g^2_3 \Big( 1 - \frac{g^2_3}{24\pi m_D} \Big), \\
\bar{g}'^2_3 = & g'^2_3, \\
\bar{\mu}^2_{11,3} =& \mu^2_{11,3} - \frac{1}{4\pi}\Big(3 h_{1} m_D +  h_{2} m_D' \Big) \nonumber \\
& \quad + \frac{1}{16\pi^2} \bigg( 3g^2_3 h_{1} - 3 h^2_{1} - {h}^2_{2} - \frac{3}{2}{h}^2_{3} \nonumber \\
& \quad + \Big(-\frac{3}{4}g^4_3 + 12 g^2_3 h_{1} \Big) \ln\Big(\frac{\Lambda_{\td}'}{2m_D} \Big) 
 - 6 h^2_{1}  \ln\Big(\frac{\Lambda_{\td}'}{2m_D + \mu_{11,3}} \Big) \nonumber \\
& \quad - 2 {h}^2_{2} \ln\Big(\frac{\Lambda_{\td}'}{2m_D' + \mu_{11,3}}
 \Big) - 3 {h}^2_{3} \ln\Big(\frac{\Lambda_{\td}'}{m_D+m_D'+ \mu_{11,3}}
 \Big) \nonumber \\
& \quad + 2 \mu_{11,3} \Big( 3\frac{h^2_{1}}{m_D} + \frac{{h}^2_{2}}{m_D'} \Big)  + 2 \mu_{22,3} \Big( 3\frac{ h_{1}h_{4}}{m_D} + \frac{  {h}_{2}{h}_{5}}{m_D'} \Big) \bigg)_{\text{2-loop}}, \\ 
\bar{\mu}^2_{22,3} =& \mu^2_{22,3} - \frac{1}{4\pi}\Big(3 h_{4} m_D +  h_{5} m_D' + 8 \omega_3 m_D'' \Big) \nonumber \\
& \quad + \frac{1}{16\pi^2} \bigg( 3g^2_3 h_{4} - 3 h^2_{4} - {h}^2_{5} - \frac{3}{2}{h}^2_{6} \nonumber \\
& \quad + \Big(-\frac{3}{4}g^4_3 + 12 g^2_3 h_{4} \Big) \ln\Big(\frac{\Lambda_{\td}'}{2m_D} \Big) 
 - 6 h^2_{4}  \ln\Big(\frac{\Lambda_{\td}'}{2m_D + \mu_{22,3}} \Big) \nonumber \\
& \quad - 2 {h}^2_{5} \ln\Big(\frac{\Lambda_{\td}'}{2m_D' +
   \mu_{22,3}} \Big) - 3 {h}^2_{6}
 \ln\Big(\frac{\Lambda_{\td}'}{m_D+m_D'+ \mu_{22,3}} \Big) \nonumber \\
& \quad + 2 \mu_{22,3} \Big( 3\frac{h^2_{4}}{m_D} + \frac{{h}^2_{5}}{m_D'} \Big)  + 2 \mu_{11,3} \Big( 3\frac{ h_{4}h_{1}}{m_D} + \frac{  {h}_{5}{h}_{2}}{m_D'} \Big)  \bigg)_{\text{2-loop}}, \\ 
\bar{\mu}^2_{12,3} =& \mu^2_{12,3}, \\
\bar{\lambda}_{1,3} =& \lambda_{1,3} - \frac{1}{8\pi}\Big( \frac{3 h^2_{1}}{m_D} + \frac{ h_{2}^2 }{m_D'} + \frac{ h_{3}^2}{m_D+m_D'} \Big), \\
\bar{\lambda}_{2,3} =& \lambda_{2,3} - \frac{1}{8\pi}\Big( \frac{3 h^2_{4}}{m_D} + \frac{ h_{5}^2 }{m_D'} + \frac{ h_{6}^2}{m_D+m_D'} \Big), \\
\bar{\lambda}_{3,3} =& \lambda_{3,3} - \frac{1}{4\pi}\Big( \frac{3 h_{1}h_{4}}{m_D} + \frac{ h_{2} h_{5} }{m_D'} + \frac{ h_{3} h_{6}}{m_D+m_D'} \Big), \\
\bar{\lambda}_{4,3} =& \lambda_{4,3} \\
\text{and} \quad \bar{\lambda}_{5,3} =& \lambda_{5,3}.
\end{align}
The temporal scalars have only a small 
effect on the running of mass parameters. We therefore fix the
renormalisation group 
scale in the resulting effective theory as
$\Lambda_{\td}' = \Lambda_{\td}$.

When the second doublet is integrated out as a heavy field, the
parameters of the final 3D theory read
\begin{align}
\hat{g}^2_3 =& {\bar{g}}^2_3 \Big( 1 - \frac{{\bar{g}}^2_3}{48\pi \widetilde{\mu}_\theta} \Big), \\
\hat{g}'^2_3 =& \bar{g}'^2_3 \Big( 1 - \frac{\bar{g}'^2_3}{48\pi \widetilde{\mu}_\theta} \Big), \\
\hat{\lambda} =& \widetilde{\lambda}_1 - \frac{1}{16\pi}\frac{1}{\widetilde{\mu}_\theta}\Big(2 \widetilde{\lambda}^2_3 + 2 \widetilde{\lambda}_3\widetilde{\lambda}_4 + \widetilde{\lambda}^2_4 + |\widetilde{\lambda}_5|^2 -48 \RE(\widetilde{\lambda}_6\widetilde{\lambda}_7 ) + 48 |\widetilde{\lambda}_6|^2 \Big), \\
\hat{\mu}^2 =& \widetilde{\mu}^2_\phi - \frac{\widetilde{\mu}_\theta}{4\pi}\Big(2 \widetilde{\lambda}_3 + \widetilde{\lambda}_4\Big) +\frac{1}{16\pi^2} \bigg(\frac{1}{8}(3\bar{g}^2_3 + \bar{g}'^2_3 )(2\widetilde{\lambda}_3 + \widetilde{\lambda}_4) - \widetilde{\lambda}^2_3 - \widetilde{\lambda}_3 \widetilde{\lambda}_4 -\widetilde{\lambda}^2_4 \nonumber \\
&  \quad + 3 \widetilde{\lambda}_2 (2\widetilde{\lambda}_3+\widetilde{\lambda}_4) + 18 \RE(\widetilde{\lambda}_7 \widetilde{\lambda}_6) -3|\widetilde{\lambda}_5|^2 \Big(\ln\Big(\frac{\Lambda_{\td}''}{2 \widetilde{\mu}_\theta}\Big) + \frac{1}{2} \Big)-3|\widetilde{\lambda}_7|^2 \Big(\ln\Big(\frac{\Lambda_{\td}''}{3 \widetilde{\mu}_\theta}\Big) + 2 \Big) \nonumber \\
& - \quad 9|\widetilde{\lambda}_6|^2 \Big( \ln\Big(\frac{\Lambda_{\td}''}{ \widetilde{\mu}_\theta}\Big) + \frac{1}{2} \Big) + \frac{1}{8} \Big(-3\bar{g}^4_3 - \bar{g}'^4_3 + 4(3\bar{g}^2_3 + \bar{g}'^2_3)(2\widetilde{\lambda}_3 + \widetilde{\lambda}_4) \nonumber \\
& \quad - 16 (\widetilde{\lambda}^2_3 + \widetilde{\lambda}_3
\widetilde{\lambda}_4 + \widetilde{\lambda}^2_4 ) \Big) \ln \Big( \frac{\Lambda_{\td}''}{2 \widetilde{\mu}_\theta} \Big)   \bigg)_{\text{2-loop}}.
\end{align}

An estimate of the validity of this last step of dimensional reduction is obtained
by deriving the dimension-six operator
$\hat{\Lambda}_6 (\phi^\dagger \phi)^3_{\td}$ omitted from our phase 
transition analysis. The dominant contribution to it originates from a tree-level diagram proportional to 
$|\widetilde{\lambda}^2_6|$~\cite{Laine:1996ms}, which we calculate at zero momentum. 
Performing parameter matching, we obtain
\begin{align}
\hat{\Lambda}_6 = \frac83 \frac{|\widetilde{\lambda}^2_6|}{m^2_\theta},
\end{align}
and computing the effective potential with this operator included gives an estimate on 
its effect on the dynamics of the effective theory. Details of the calculation are
to be found in the companion paper~\cite{Gorda:2018hvi}.

The renormalisation scale of the final 3D theory is fixed at the light scale
as $\Lambda_{\td}'' = \hat{g}^2_3$ to match the choice of Ref.~\cite{Kajantie:1995kf}.
In regions of the parameter space where the transition is of first order, $\hat{\lambda}$ 
is necessarily small. Consequently, the effects of running are very mild in the final effective theory.
In particular, we have verified that the six-dimensional estimate of Fig.~\ref{fig:Veff3d} is
not sensitive to the choice of $\Lambda_{\td}''$.

\end{widetext}

\end{document}